\documentstyle[twocolumn,prl,aps,floats,psfig]{revtex}
\begin{document}
\draft

\title{Accurate Charge-Dependent Nucleon-Nucleon Potential
at Fourth Order of Chiral Perturbation Theory}

\author{D. R. Entem$^{1,2,}$\thanks{Electronic addresses: 
        dentem@uidaho.edu, entem@usal.es}
        and R. Machleidt$^{1,}$\thanks{Electronic address: machleid@uidaho.edu}}

\address{$^1$Department of Physics, University of Idaho, Moscow, ID 83844, USA\\
  $^2$Nuclear Physics Group, University of Salamanca, E-37008 Salamanca, Spain}

\date{\today}

\maketitle

\begin{abstract}
We present the first nucleon-nucleon potential at
next-to-next-to-next-to-leading order (fourth order)
of chiral perturbation theory.
Charge-dependence is included up to next-to-leading order
of the isospin-violation scheme.
The accuracy for the reproduction of the $NN$ data below 290 MeV
lab.\ energy is comparable to the one of
phenomenological high-precision potentials.
Since $NN$ potentials of order three and less are known to be
deficient in quantitative terms, the present work shows that the
fourth order is necessary and sufficient for a
reliable $NN$ potential derived from chiral
effective Lagrangians.
The new potential provides a promising starting point
for exact few-body calculations and microscopic nuclear
structure theory (including chiral many-body forces
derived on the same footing).
\end{abstract}


The theory of nuclear forces has a long history.
Based upon the Yukawa idea~\cite{Yuk35}, first field-theoretic
attempts~\cite{TMO52,BW53}
to derive the nucleon-nucleon ($NN$) interaction
focused on pion-exchange,
resulting in the $NN$ potentials
by Gartenhaus~\cite{Gar55} and
by Signell and Marshak~\cite{SM58}. 
However, even qualitatively, these potentials barely
agreed with empirical information on the
nuclear force.
So, these ``pion theories'' of the 1950s
are generally judged as failures---for reasons
we understand today: pion dynamics is constrained by chiral
symmetry, a crucial point that was unknown in the 1950s.

Historically, the experimental discovery of heavy 
mesons~\cite{Erw61} in the early 1960s
saved the situation. The one-boson-exchange (OBE)
model~\cite{OBEP,Mac89} emerged which is still the most economical
and quantitative
phenomenology for describing the 
nuclear force~\cite{Sto94,Mac01}.
The weak point of this model, however, is the scalar-isoscalar
``sigma'' or ``epsilon'' boson, for which the empirical
evidence remains controversial. Since this boson is associated
with the  correlated (or resonant) exchange of two pions,
a vast theoretical effort that occupied more than a decade 
was launched to derive the 2$\pi$-exchange contribution
of the nuclear force, which creates the intermediate 
range attraction.
For this, dispersion theory as well as 
field theory were invoked producing  the
Paris~\cite{Vin79,Lac80} and the Bonn~\cite{MHE87,Mac89}
potentials.

The nuclear force problem appeared to be solved; however,
with the discovery of quantum chromo-dynamics (QCD), 
all ``meson theories'' had to
be relegated to models and the attempts to derive
the nuclear force started all over again.

The problem with a derivation from QCD is that
this theory is non-perturbative in the low-energy regime
characteristic of nuclear physics, which makes direct solutions
impossible.
Therefore, during the first round of new attempts,
QCD-inspired quark models~\cite{MW88} became popular. 
These models were able to reproduce
qualitatively some of the gross features of the nuclear force.
However, on a critical note, it has been pointed out
that these quark-based
approaches were nothing but
another set of models and, thus, did not represent any
fundamental progress. Equally well, one may then stay
with the simpler and much more quantitative meson models.

A major breakthrough occurred when 
the concept of an effective field theory (EFT) was introduced
and applied to low-energy QCD.
As outlined by Weinberg in a seminal paper~\cite{Wei79},
one has to write down the most general Lagrangian consistent
with the assumed symmetry principles, particularly
the (broken) chiral symmetry of QCD.
At low energy, the effective degrees of freedom are pions and
nucleons rather than quarks and gluons; heavy mesons and
nucleon resonances are ``integrated out''.
So, in a certain sense we are back to the 1950s,
except that we are smarter by 40 years of experience:
broken chiral symmetry is a crucial constraint that generates
and controls the dynamics and establishes a clear connection
with the underlying theory, QCD.

The chiral effective Lagrangian is given by an infinite series
of terms with increasing number of derivatives and/or nucleon
fields, with the dependence of each term on the pion field
prescribed by the rules of broken chiral symmetry~\cite{Wei90}.
Applying this Lagrangian to $NN$ scattering generates an unlimited
number of Feynman diagrams, which may suggest
again an untractable problem.
However, Weinberg showed~\cite{Wei90} that a systematic expansion
of the nuclear amplitude exists in terms of $(Q/\Lambda_\chi)^\nu$,
where $Q$ denotes a momentum or pion mass, 
$\Lambda_\chi \approx 1$ GeV is the chiral symmetry breaking
scale, and $\nu \geq 0$.
For a given order $\nu$, the number of contributing terms is
finite and calculable; these terms are uniquely defined and
the prediction at each order is model-independent.
By going to higher orders, the amplitude can be calculated
to any desired accuracy.
The scheme just outlined has become known as chiral perturbation
theory ($\chi$PT).

Following the first initiative by Weinberg \cite{Wei90}, pioneering
work was performed by Ord\'o\~nez, Ray, and
van Kolck \cite{ORK94,Kol99} who 
constructed a $NN$ potenial in coordinate space
based upon $\chi$PT at
next-to-next-to-leading order (NNLO; $\nu=3$).
The results were encouraging and
many researchers~\cite{CPS92} became attracted to the new field.
Kaiser, Brockmann, and Weise~\cite{KBW97} presented the first model-independent
prediction for the $NN$ amplitudes of peripheral
partial waves at NNLO.
Epelbaum {\it et al.}~\cite{EGM98} developed the first momentum-space
$NN$ potential at NNLO.

\begin{table}
\caption{Low-energy constants applied in the N$^3$LO $NN$ potential 
(column `$NN$').
The $c_i$ belong to the dimension-two $\pi N$ Lagrangian
and are in units of GeV$^{-1}$, while the 
$\bar{d}_i$ are associated with the dimension-three Lagrangian
and are in units of GeV$^{-2}$. The column `$\pi N$' shows values
determined from $\pi N$ data.}
\begin{tabular}{cdc}
 & $NN$ & $\pi N$ \\
\hline
$c_1$ & --0.81 & $-0.81\pm 0.15^a$ \\
$c_2$ & 2.80 & $3.28\pm 0.23^b$ \\
$c_3$ & --3.20 & $-4.69\pm 1.34^a$ \\
$c_4$ & 5.40 & $3.40\pm 0.04^a$ \\
$\bar{d}_1 + \bar{d}_2$ & 3.06 & $3.06\pm 0.21^b$ \\
$\bar{d}_3$ & --3.27 & $-3.27\pm 0.73^b$ \\
$\bar{d}_5$ & 0.45 & $0.45\pm 0.42^b$ \\
$\bar{d}_{14} - \bar{d}_{15}$ & --5.65 & $-5.65\pm 0.41^b$
\end{tabular}
\footnotesize
$^a$Table~1, Fit~1 of Ref.~\cite{BM00}.\hspace{5mm}
$^b$Table~2, Fit~1 of Ref.~\cite{FMS98}.
\end{table}

In the 1990s, unrelated, parallel research showed
that, for conclusive few-body calculations and meaningful microscopic nuclear
structure predictions, the input $NN$ potential must be
of the highest precision; i.~e., it must reproduce the $NN$ data below
about 300 MeV lab.\ energy with a $\chi^2/$datum $\approx 1$.
The family of high-precision $NN$ potentials~\cite{Sto94,WSS95,MSS96,Mac01} was
developed which fulfills this requirement.
Due to the outstanding accuracy of these $NN$ potentials,
it was possible to pin down cases of few-body scattering
and of nuclear structure that clearly 
require three-nucleon forces (3NF) for their miscroscopic explanation.
Famous examples are the $A_y$ puzzle of $N$-$d$ scattering~\cite{Glo96}
and the ground state of $^{10}$B~\cite{Cau02}.

One important advantage of $\chi$PT is that it makes specific
predictions for many-body forces. For a given order of $\chi$PT,
both 2N and 3N forces are generated on the same footing. 
At next-to-leading order (NLO),
all 3NF cancel~\cite{Wei90,Kol94}; 
however, at NNLO and higher orders, well-defined, nonvanishing 3NF terms occur.
As discussed, since 3NF effects are in general very subtle,
it is only possible to demonstrate their necessity and relevance
when the 2NF is of high precision.

$NN$ potentials based upon 
$\chi$PT at NNLO are poor in quantitative terms; they
reproduce the $NN$ data below 290 MeV
lab.\ energy with a $\chi^2$/datum of more than 20 which is
totally unacceptable. Clearly, there is a strong need for more precision,
implying that going to higher order is necessary.

It is the purpose of this note to present the first $NN$
potential that is based consistently on $\chi$PT at
next-to-next-to-next-to-leading order (N$^3$LO; fourth order). 
We will show that, at this order, 
the accuracy is comparable to the one of the
high-precision phenomenological potentials. 
Thus, the $NN$ potential at N$^3$LO 
is the first to meet the requirements for a reliable 
input-potential for exact few-body
and microscopic nuclear structure calculations (including chiral 3NF
consistent with the chiral 2NF).

\begin{table}
\caption{
$\chi^2$/datum for the reproduction of the 1999 $np$ 
database~\protect\cite{note2} below 290 MeV by various $np$ potentials.}
\begin{tabular}{cccccr}
 Bin (MeV) 
 & \# of data 
 & N$^3$LO$^a$
 & NNLO$^b$
 & NLO$^b$ 
 & AV18$^c$
\\
\hline 
0--100&1058&1.06&1.71&5.20&0.95\\
100--190&501&1.08&12.9&49.3&1.10\\
190--290&843&1.15&19.2&68.3&1.11\\
\hline
0--290&2402&1.10&10.1&36.2&1.04
\end{tabular}
\footnotesize
$^a$This work. \hspace{5mm}
$^b$Ref.~\cite{Epe02}. \hspace{5mm}
$^c$Ref.~\cite{WSS95}.

\end{table}

\begin{table}[b]
\caption{
$\chi^2$/datum for the reproduction of the 1999 $pp$ 
database~\protect\cite{note2} below 290 MeV by various $pp$ potentials.}
\begin{tabular}{cccccr}
 Bin (MeV) 
 & \# of data
 & N$^3$LO$^a$
 & NNLO$^b$
 & NLO$^b$
 & AV18$^c$
\\
\hline 
0--100&795&1.05&6.66&57.8&0.96\\
100--190&411&1.50&28.3&62.0&1.31\\
190--290&851&1.93&66.8&111.6&1.82\\
\hline
0--290&2057&1.50&35.4&80.1&1.38\\
\end{tabular}
\footnotesize
$^a$This work. \hspace{5mm}
$^b$See footnote~\cite{note3}. \hspace{5mm}
$^c$Ref.~\cite{WSS95}.\\
\end{table}

In $\chi$PT, the $NN$ amplitude is uniquely determined
by two classes of contributions: contact terms and pion-exchange
diagrams. At N$^3$LO, there are two contacts of order $Q^0$ 
[${\cal O}(Q^0)$], 
seven of ${\cal O}(Q^2)$, 
and 15 of ${\cal O}(Q^4)$, 
resulting in a total of 24 contact terms, which generate 24 
parameters that are crucial for the fit of the partial waves
with orbital angular momentum $L\leq 2$~\cite{note1}.

Now, turning to the pion contributions:
At leading order [LO, ${\cal O}(Q^0)$, $\nu=0$], 
there is only the wellknown static one-pion exchange (OPE). 
Two-pion exchange (TPE) starts
at next-to-leading order (NLO, $\nu=2$), and there are
further TPE
contributions in any higher order.
While TPE at NNLO was known for a while~\cite{ORK94,KBW97,EGM98},
TPE at N$^3$LO has been calculated only recently by
Kaiser~\cite{Kai01}. All $2\pi$ exchange contributions up to
N$^3$LO are summarized in a pedagogical and systematic
fashion in Ref.~\cite{EM02} where 
the model-independent results for $NN$ scattering in peripheral
partial waves are also shown. We use the analytic expressions
published in Ref.~\cite{EM02}.
Finally, there is also three-pion exchange, which
shows up for the first time 
at N$^3$LO (two loops). 
In Ref.~\cite{Kai99},
it was demonstrated that the 3$\pi$ contributions at this order
are negligible, which is why we leave them out.

For an accurate fit of the low-energy $pp$ and $np$ data, 
charge-dependence is important.
We include charge-dependence up to next-to-leading order 
of the isospin-violation scheme 
(NL\O, in the notation of Ref.~\cite{WME01}).
Thus, we include
the pion mass difference in OPE and the Coulomb potential
in $pp$ scattering, which takes care of the L\O\/ contributions. 
At order NL\O\, we have pion mass difference in the NLO part of TPE,
$\pi\gamma$ exchange~\cite{Kol98}, and two charge-dependent
contact interactions of order $Q^0$ which make possible
an accurate fit of the three different $^1S_0$ scattering 
lengths, $a_{pp}$, $a_{nn}$, and $a_{np}$.

\begin{figure}
\vspace{-1.6cm}
\hspace{-1.5cm}
\psfig{figure=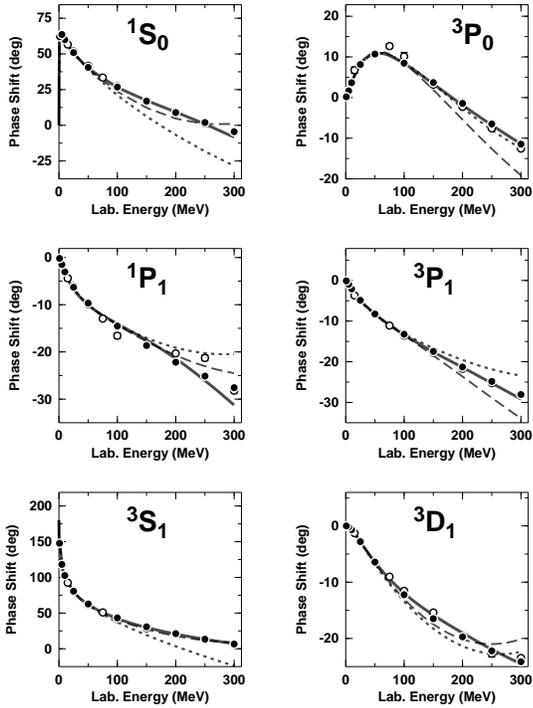,height=13cm}
\vspace{-1.9cm}
\caption{$np$ phase parameters below 300 MeV lab.\ energy
for partial waves with $J\leq 2$. 
The solid line is the result at N$^3$LO. 
The dotted and dashed lines are the phase shifts at NLO and NNLO,
respectively, as obtained by Epelbaum {\it et al.} \protect\cite{Epe02}.
The solid dots show the Nijmegen
multi-energy $np$ phase shift analysis~\protect\cite{Sto93}, and 
the open circles are the VPI
single-energy $np$ analysis SM99~\protect\cite{SM99}.}
\end{figure}

Chiral perturbation theory is a low-momentum expansion.
It is valid only for momenta $Q \ll \Lambda_\chi \approx 1$ GeV.
To enforce this, we multiply all expressions (contacts and
irreducible pion exchanges) with a regulator function,
\begin{equation}
\exp\left[ 
-\left(\frac{p}{\Lambda}\right)^{2n}
-\left(\frac{p'}{\Lambda}\right)^{2n}
\right] \; ,
\end{equation}
where $p$ and $p'$ denote, respectively, the magnitudes
of the initial and final nucleon momenta in the center-of-mass
frame. We use $\Lambda = 0.5$ GeV throughout. The exponent $2n$ is chosen
to be sufficiently large so that 
the regulator generates powers which are beyond
the order ($\nu=4$) at which our calculation is conducted; i.~e.,
terms up to $Q^4$ are not affected.

\begin{figure}
\vspace{-1.6cm}
\hspace{-1.5cm}
\psfig{figure=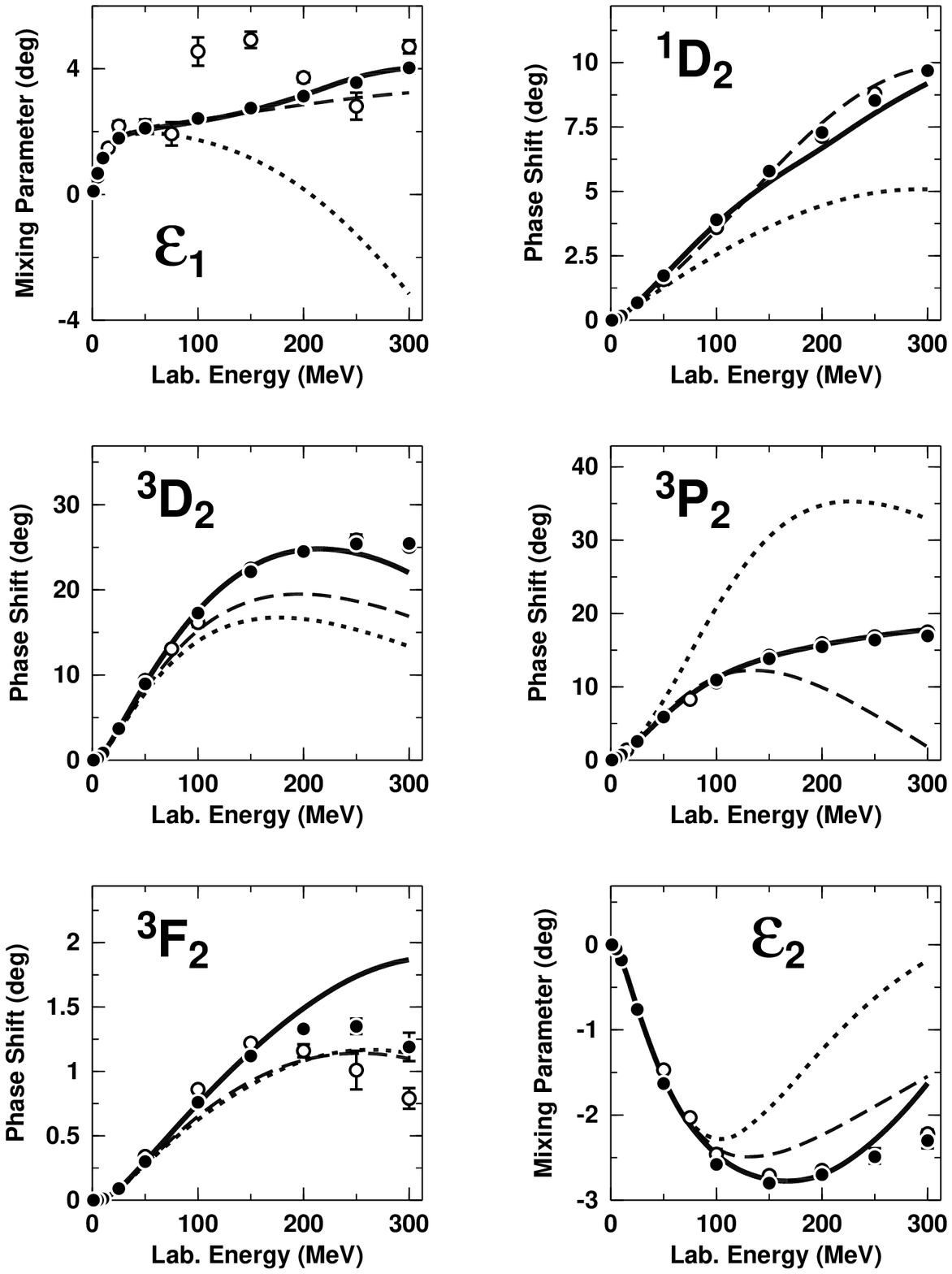,height=13cm}
\vspace{-1.6cm}
\parbox{7cm}{
\footnotesize
\vspace{-2.9cm} 
\hspace{2.6cm}
FIG.~1, continued.}
\end{figure}

The contact terms plus irreducible pion-exchange expressions at N$^3$LO,
multiplied by the above regulator, define the $NN$ potential at N$^3$LO.
This potential is applied in a Lippmann-Schwinger equation to obtain
the $T$-matrix from which phase shifts and $NN$ 
observables are calculated. The corresponding homogenous equation determines
the properties of the two-nucleon bound state (deuteron).

The peripheral partial waves of $NN$ scattering with $L\geq 3$
are exclusively determined by OPE and TPE because
the N$^3$LO contacts contribute to $L\leq 2$ only.
OPE and TPE at N$^3$LO depend on
the axial-vector coupling constant, $g_A$ (we use $g_A=1.29$),
the pion decay constant, $f_\pi=92.4$ MeV,
and eight low-energy constants (LEC) that appear in the dimension-two 
and dimension-three $\pi N$ Lagrangians (cf.\ Ref.~\cite{EM02}).
In the optimization process,
we varied three of them, namely, $c_2$, $c_3$, and $c_4$.
We found that the other LEC are not very effective in the
$NN$ system and, therefore, we kept them at the
values determined from $\pi N$ (cf.\ Table~I).
The most influential constant is $c_3$, which has to be chosen
on the low side (slightly more than one standard deviation
below its $\pi N$ determination) for an optimal fit of the $NN$ data. 
Our choice for $c_4$, which is substantially above
the value determined in $\pi N$,
is necessary to bring the $^3F_2$ phase shift down.

The most important set of fit parameters are the ones associated
with the 24 contact terms that rule the partial waves
with $L\leq 2$. In addition, we have two charge-dependent
contacts, which brings the number of contact parameters
to 26. Since we treated three LEC as semi-free,
the total number of parameters of the N$^3$LO potential is 29.

In the optimization procedure, we fit first phase shifts,
and then we refine the fit by minimizing the
$\chi^2$ obtained from a direct comparison with the data.
The phase shifts at N$^3$LO for $np$ scattering below 300 MeV
lab.\ energy are displayed in Fig.~1.
The $\chi^2/$datum for the fit of the $np$ data below
290 MeV is shown in Table~II, and the corresponding one for $pp$
is given in Table~III.
The $\chi^2$ tables demonstrate a dramatic improvement
of the $NN$ interaction order by order.
It is clearly revealed that, at NLO and NNLO,
the reproduction of the $NN$ data is of unacceptably
poor quality.
However, at N$^3$LO, the quantitative character is comparable
to  the phenomenological high-precision Argonne $V_{18}$
potential~\cite{WSS95}.

In conclusion, we have developed the first $NN$ potential
at fourth order of $\chi$PT~\cite{note4}. 
This potential is as quantitative
as some so-called high-precision phenomenological potentials.
Due to its basis in $\chi$PT,
the many-body forces associated with this two-body force
are well-defined. 
Thus, we have a promising starting point
for exact few-body calculations and microscopic nuclear
structure theory.

This work was supported by the U.S. National Science
Foundation under Grant No.~PHY-0099444 and by three Spanish foundations:
the Ministerio de Ciencia y Tecnolog{\'\i}a 
under Contract No. BFM2001-3563, the Junta de Castilla y Le\'on under 
Contract No. SA-109/01, and the Ram\'on Areces Foundation.


\begin{references}
\bibitem{note4} A user-friendly computer code for the N$^3$LO
potential is available from the authors upon request.
\end{references}
\end{document}